\documentclass[journal]{IEEEtran}

\usepackage{multicol}
\usepackage{graphicx}
\graphicspath{{Figures/}}
\usepackage{epstopdf}
\usepackage{acronym}
\usepackage{amsmath}
\usepackage{color, colortbl}
\usepackage{subfigure}
\usepackage[english]{babel}
\usepackage{blindtext}
\usepackage{algorithm} 
\usepackage{algorithmic} 
\usepackage{multirow}
\usepackage{color}
\usepackage{subfigure}
\usepackage{balance}
\usepackage{caption}

\begin{document}

\newacro{5G}{5{th} generation}

\newacro{BER}{bit error rate}

\newacro{PAPR}{peak-to-average power ratio}

\newacro{SEFDM}{spectrally efficient frequency division multiplexing}

\title{Index Modulation Pattern Design for Non-Orthogonal Multicarrier Signal Waveforms}

\author{{Yinglin Chen, Tongyang Xu,~\IEEEmembership{Member,~IEEE} and Izzat Darwazeh,~\IEEEmembership{Senior Member,~IEEE}}
\thanks{
This work was supported in part by the Engineering and Physical Sciences Research Council ``Impact Acceleration Discovery to Use'' Award under Grant EP/R511638/1; and in part by Cisco University Research Program Fund, which funded our programme of research entitled: Non-Orthogonal IoT for Future Wireless Networks and 5G.

Y. Chen is with the 6G Research Center, China Telecom Research Institute, Guangzhou 510660, China (e-mail: chenyl37@chinatelecom.cn). 

T. Xu is with the School of Engineering, Newcastle University, Newcastle upon Tyne, NE1 7RU, U.K., and also with the Department of Electronic and Electrical Engineering, University College London (UCL), London, WC1E 7JE, U.K. (e-mail: tongyang.xu@ieee.org) 

I. Darwazeh is with the Department of Electronic and Electrical Engineering, University College London (UCL), London, WC1E 7JE, U.K. (e-mail: i.darwazeh@ucl.ac.uk).

}}

\maketitle

\begin{abstract}

Spectral efficiency improvement is a key focus in most wireless communication systems and achieved by various means such as using large antenna arrays and/or advanced modulation schemes and signal formats. This work proposes to further improve spectral efficiency through combining non-orthogonal spectrally efficient frequency division multiplexing (SEFDM) systems with index modulation (IM), which can efficiently make use of the indices of activated subcarriers as communication information. Recent research has verified that IM may be used with SEFDM to alleviate inter-carrier interference (ICI) and improve error performance. This work proposes new SEFDM signal formats based on novel activation pattern designs, which limit the locations of activated subcarriers and enable a variable number of activated subcarriers in each SEFDM subblock. SEFDM-IM system designs are developed by jointly considering activation patterns, modulation schemes and signal waveform formats, with a set of solutions evaluated under different spectral efficiency scenarios. Detailed modelling of coded systems and simulation studies reveal that the proposed designs not only lead to better bit error rate (BER) but also lower peak-to-average power ratio (PAPR) and reduced computational complexity relative to other reported index-modulated systems.

\end{abstract}

\begin{IEEEkeywords}
Index modulation, waveform, multi-carrier, subcarrier pattern, inter-carrier interference, OFDM, SEFDM, non-orthogonal, spectral efficiency, PAPR. 
\end{IEEEkeywords}

\section{Introduction}

\IEEEPARstart{I}{ndex} modulation (IM) is emerging as a new waveform format due to the need for increasing spectral efficiency and energy efficiency \cite{IM_nextGeneration,IMfor5G}. By activating a subset of communication resource entities during each signalling period, implicit information is embedded in activation patterns and transmitted without energy consumption. Meanwhile, activated entities operate with conventional digital modulation schemes and carry explicit information. According to selected resource entities, IM is classified into three main domains and termed spatial-domain IM (SD-IM), time-domain IM (TD-IM), and frequency-domain IM (FD-IM). 

Spatial modulation (SM), proposed in \cite{spatial_modulation_2008}, was an early implementation of SD-IM, where only a single transmit (TX) antenna in a multiple-input multiple-output (MIMO) system is activated in each symbol period. Inter-channel interference is therefore removed because of one active TX antenna, which leads to reduced detection complexity but at the price of throughput loss. Subsequent research activated multiple TX antennas to compensate for the throughput loss such as generalized-SM (GSM) \cite{GSM} and multiple active-SM (MA-SM) \cite{MA_SM}, in which GSM transmits the same data symbols on all activated antennas to avoid inter-channel interference and inter-symbol interference (ISI) while MA-SM transmits different symbols on different antennas. TX antenna grouping was introduced to GSM to harvest transmission diversity \cite{SM_TX_grouping}.
Likewise, SD-IM can also be performed at receive (RX) antennas. Pre-processing aided-SM (PSM) deploys pre-coding to ensure that only the target RX antenna receives signals while the others only receive noise. In addition, the index of the activated RX antenna conveys information \cite{PSM}. Later, the SD-IM scheme was generalized in \cite{GPSM} to allow multiple activated RX antennas. In addition to the aforementioned SM schemes that utilize the indices of antennas to transmit information, MIMO with Antenna Number Modulation (MIMO-ANM) was developed in \cite{MIMO-ANM} which utilizes the number of antennas instead of their indices. By enabling a variable number of activated TX antennas, MIMO-ANM exhibits performance advantage over SM.

TD-IM is performed in time domain where only a subset of time slots is activated for data transmission. Alternatively, empty time slots can be utilized by using a different modulation scheme that is distinguishable from the one modulated at activated time slots. When such TD-IM techniques are applied on single-carrier systems, they are termed either single-carrier IM (SC-IM) \cite{Single_carrier_TD_IM} or dual-mode SC-IM (DM-SCIM) \cite{dual_mode_SC_TD_IM}. To further enhance achievable spectral efficiency, single-mode and dual-mode TD-IM are combined with Faster-than-Nyquist (FTN), which enhances spectral efficiency by violating the time-orthogonal Nyquist-criterion \cite{dual_mode_SC_TD_IM, single_mode_FTN_TD_IM}. Particularly, FTN-IM in \cite{single_mode_FTN_TD_IM} can efficiently mitigate ISI owing to the unused time slots. In addition to the aforementioned TD-IM schemes that purely work in time domain, TD-IM is also incorporated with SD-IM leading to a family of multidimensional IM schemes, where index information is transmitted through the indices of activated time slots and TX antennas. There has been substantial research in this area since the early implementation of space-time IM (ST-IM) \cite{space_time_IM}. In the recent work \cite{FTN_multimode_ST_IM}, FTN in multi-mode TD-IM is combined with SM, which yields better BER performance than traditional MIMO.

FD-IM refers to IM performed on subcarriers' state in multi-carrier systems. Orthogonal frequency division multiplexing with index modulation (OFDM-IM) proposed in \cite{OFDM_IM_Basar} paves the way for subsequent research by introducing subcarrier grouping. In this case, a fixed number of subcarriers are activated in each subcarrier group (i.e., subblock) to simplify index selection and detection. As index information is conveyed implicitly without energy consumption, signal power originally allocated to unused subcarriers can be suppressed, which yields improved energy efficiency. Advantageously, the saved signal power can be allocated to activated subcarriers, leading to improved error performance over OFDM. In addition, peak-to-average power ratio (PAPR) of index-modulated multi-carrier signals is reduced, given that the number of activated subcarriers is reduced \cite{OFDMIM_PAPR}. Hence, the energy efficiency of FD-IM systems is improved. To obtain higher achievable spectral efficiency, enhanced OFDM-IM schemes have been put forward by increasing the flexibility of activation patterns. In \cite{Generalized_OFDM_IM}, OFDM with generalized index modulation 1 (OFDM-GIM1) with a variable number of activated subcarriers per subblock and OFDM-GIM2 with indexing on the in-phase and quadrature components of subcarriers were proposed. 
On the other hand, OFDM with subcarrier number modulation (OFDM-SNM) \cite{OFDM_Subcarrier_Number_Modulation,enahnced_OFDM_Subcarrier_Number_Modulation} was developed to perform IM on the number of activated subcarriers instead of their indices, which obtains additional coding gain by adjusting activated subcarriers' locations based on channel conditions. 
However, spectral efficiency of OFDM-SNM is variable since it is dependant on the number of activated subcarriers, which may lead to error propagation. To solve this problem, joint-mapping OFDM-SNM (JM-OFDM-SNM) was put forward which ensures a fixed length of information bits for each transmission by jointly considering subcarrier activation patterns and modulation schemes, and performance improvement was reported \cite{joint-mapping-OFDM-SNM}. Considering the throughput loss brought by the unused subcarriers, a group of OFDM-IM schemes that have all subcarriers activated were proposed. In dual-mode OFDM (DM-OFDM) \cite{dual_mode_OFDM_IM} and generalized DM-OFDM (GDM-OFDM) \cite{Generalized_Dual_Mode_OFDM_IM}, subcarriers are modulated with two distinguishable modulation schemes, and their indices transmit index information. Moreover, multiple-mode OFDM-IM (MM-OFDM-IM) \cite{MM-OFDM-IM} and its generalized version (GMM-OFDM-IM) \cite{generalized-MM-OFDM-IM} were proposed, where the permutations of multiple distinguishable modulation schemes serve IM purpose. These schemes fully utilize frequency resources by trading off energy efficiency and detection complexity.

OFDM is a key constituent of many modern wireless systems in 4G \cite{LTE_standard}, 5G \cite{5gtutorial} and wireless local area network (WLAN) \cite{WLAN_standard}. Separated by a certain spacing, subcarriers overlap with each other without experiencing inter-carrier interference (ICI), which enables low-complexity receiver designs. However, OFDM is constrained by such spacing to maintain the orthogonality between subcarriers. Facing the ever-increasing bandwidth demand, non-orthogonal multi-carrier systems are gaining research interests. Spectrally efficient frequency division multiplexing (SEFDM) was developed enabling flexible subcarrier spacing \cite{SEFDM2003}. Despite enhanced spectral efficiency, SEFDM suffers from self-introduced ICI between subcarriers, which degrades the performance of linear detection techniques such as zero forcing (ZF) and minimum mean square error (MMSE). Hence, complicated receiver designs are required for signal detection such as maximum likelihood (ML) detection \cite{SEFDM_MMSE_ML}, turbo equalizer \cite{softdetector, TongyangTVT2017} and deep learning algorithms \cite{deeplearning}, which limit the practical implementation of SEFDM.

To further improve spectral efficiency, IM was introduced to SEFDM \cite{SEFDMIM_CHINA, SEFDMIM_JAPAN,SEFDM-IM_jointChannel}, which effectively reduces ICI effect by switching off some subcarriers. Nevertheless, residual ICI still leads to the requirement of complicated detection techniques. As always, ML detection yields optimal performance, but it is only computationally practical in small-size systems with a small number of subcarriers and low modulation cardinality.  
In \cite{SEFDMIM_CHINA}, IM was applied in SEFDM by activating only a subset of SEFDM subcarriers in each subblock, and subblock-based ML detection was implemented. To reduce the computational complexity from ML, log-likelihood ratio (LLR) detection has been investigated. A successive MMSE-LLR receiver was developed in \cite{SEFDMIM_JAPAN}, where an MMSE equalizer was deployed for ICI mitigation, followed by an LLR detector to estimate subcarriers’ state. Channel coding was reported for several traditional SEFDM systems demonstrating advantages of low-density  parity-check (LDPC) coding in simulation research \cite{Xinyue_6G_Coexistence} and practical experiment \cite{Hedaia_TMTT_2019}. For SEFDM-IM, LDPC was initially used in \cite{SEFDM-IM_jointChannel}, where improved bit error rate (BER) was achieved with the aid of turbo equalization after sufficient iterations. In addition, channel estimation methods  were put forward in \cite{SEFDM-IM_jointChannel} for SEFDM-IM systems. However, we find that all previous research on SEFDM-IM deployed the same IM design as the classical OFDM-IM in \cite{OFDM_IM_Basar}, where all subcarriers are probable to be activated. In this case, the influence of subcarrier location on the ICI level was ignored. Moreover, only BPSK and QPSK modulation were investigated.

Against this background, this work aims to propose a novel SEFDM-IM system design with enhanced ICI mitigation from an IM perspective. First, novel index modulation activation patterns for SEFDM-IM are designed to deal with inter-subblock ICI. In this case, independent signal detection per subblock will be more robust even with higher levels of subcarrier spacing compression. Moreover, different modulation schemes are jointly investigated to optimize achievable spectral efficiency. IM-based systems are normally used for low spectral efficiency applications. The proposed SEFDM-IM systems improve further spectral efficiency, which would be obviously beneficial to Internet of things (IoT) applications such as narrowband IoT (NB-IoT) and enhanced NB-IoT (eNB-IoT) \cite{Tongyang_NB_IoT_2018}. Furthermore, the enhanced index would improve the performance for visible light communications (VLC) \cite{VLC_SEFDM2016}, integrated sensing and communications (ISAC) \cite{Tongyang_ISAC_2022} and waveform-defined security (WDS) \cite{Tongyang_JIOT_WDS_2021}. For details, the main contributions of this paper are summarised as the following.

\begin{itemize}
\item{ We propose a novel pattern design principle for SEFDM-IM, which limits the locations of activated subcarriers and reduces the ICI between neighbouring subblocks. The unavailability of certain subcarrier location is compensated by the design feature of allowing systems with a variable number of activated subcarriers per subblock. Following the proposed principle, we develop three novel subcarrier activation schemes denoted as SEFDM-IM-1, SEFDM-IM-2 and SEFDM-IM-3.  }

\item{ We consider jointly waveform formats, modulation schemes, levels of bandwidth compression, and the number of activated subcarriers for four achievable spectral efficiency scenarios. To figure out an optimal index-modulated system design in each case, the system performance of three proposed schemes in coded scenarios is compared with that of the traditional SEFDM-IM and classical OFDM-IM, in terms of BER and PAPR performance, computational complexity and achievable spectral efficiency. }

\item{ This work verifies that our proposed SEFDM-IM systems can achieve better BER, lower PAPR and lower computational complexity than other available index-modulated systems when considering the same spectral efficiency. In detail, simulation results reveal that at low spectral efficiency of 0.75 bit/s/Hz, traditional SEFDM-IM outperforms OFDM-IM and our proposed SEFDM-IM systems in terms of BER and PAPR performance. When spectral efficiency is increased to 1, 1.1 and 1.25 bit/s/Hz, our proposed SEFDM-IM systems outperform OFDM-IM and traditional SEFDM-IM in both BER and PAPR performance. Results also reveal that our proposed SEFDM-IM systems achieve lower computational complexity over that of OFDM-IM and traditional SEFDM-IM for spectral efficiency of 1 bit/s/Hz and above. }

\item{ This work potentially provides a design principle for optimal SEFDM-IM. Previous work has only considered modulation order up to QPSK. We extend the modulation order up to 16QAM and observe that SEFDM-IM systems with a smaller number of activated subcarriers achieve both improved BER and PAPR performance as well as reduced computational complexity. }

\end{itemize}

The rest of this paper is organized as follows. The system model of the coded SEFDM-IM is presented in Section \ref{section1}. In Section \ref{section2}, the proposed pattern design principle and three proposed SEFDM-IM schemes are detailed. We then provide possible system configurations for four selected values of spectral efficiency in Section \ref{section3}, and the simulation results are shown in Section \ref{section4}. The computational complexity analysis is given in Section \ref{section5}. Finally, Section \ref{section6} concludes this work.

\section{System Model}\label{section1}

\begin{figure*}[ht]
    \centering
    \includegraphics[width=0.9\textwidth]{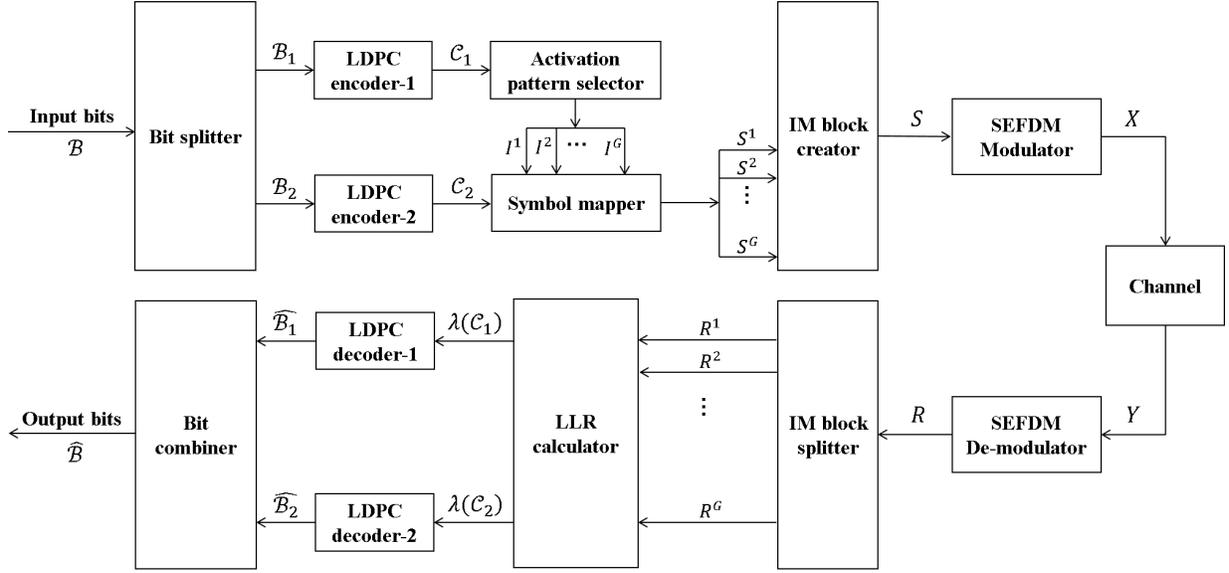}
    \caption{System diagram of the coded SEFDM-IM system.}
    \label{Fig:coded_system_disgram}
\end{figure*}

In this section, we provide a LDPC-coded SEFDM-IM system model operating in an additive white Gaussian noise (AWGN) channel. 

\subsection{Transmitter Design}

The system block diagram is shown in Fig. \ref{Fig:coded_system_disgram}. A \(B\)-length information bit sequence denoted as \({\mathcal{B}}\) is input for the transmission of one SEFDM-IM symbol, which is split into an index bit sequence \({{\mathcal{B}}_1}\) and a data bit sequence \({{\mathcal{B}}_2}\). These sequences are encoded by LDPC encoder-1 and encoder-2 of the coding rate \(\mathcal{R}\) separately, yielding two corresponding coded bit sequences \({\mathcal{C}_1}\) and \({\mathcal{C}_2}\). The coded index bit \({\mathcal{C}_1}\) is equally divided into \(G\) groups, and each group of \(L_1\) bits determines the activation pattern for an SEFDM-IM subblock of length \(K\), where \(K\) is the number of subcarriers in one subblock. The activation pattern for the \(g\)-th subblock is given by \begin{eqnarray}\label{eq:activation_pattern}
I^{g}=\left \{ i_{1}^{g},i_{2}^{g},\cdot \cdot \cdot ,i_{K}^{g} \right \},
\end{eqnarray}
where \(1\leq g\leq G\), and \(i_{\rho}^{g}\in\left \{ 0,1 \right \}\) for \(\rho =1, 2,..., K\). \(i_{\rho}^{g}=1\) denotes an activated subcarrier, and \(i_{\rho}^{g}=0\) denotes an unused subcarrier. 
In non-index-modulated systems, all \(K\) subcarriers are activated. However, in our proposed SEFDM-IM systems, only \(K_A\) out of \(K\) subcarriers are activated, and the throughput loss due to unused subcarriers is compensated by the index bits transmitted by activation patterns. The maximum number of index bits transmitted by one subblock is \(L_1=\left \lfloor \log _2\binom{K}{K_{A}} \right \rfloor\), where \(\left \lfloor \cdot \right \rfloor\) denotes the floor function. Since \(\log _2\binom{K}{K_{A}} \) may not be an integer, we use only \(U=2^{L_{1}}\) valid activation patterns and the set of which is denoted as \(\mathcal{I}\).

The coded data bit sequence \({\mathcal{C}_2}\) is equally partitioned into \(G\) groups, and each group of \(L_2=K_A\log _2M\) bits is mapped into \(K_A\) data symbols with conventional \(M\)-ary amplitude and/or phase modulation schemes. Therefore, the data symbol vector for the \(g\)-th subblock is given by
\begin{eqnarray}\label{eq:data_symbol}
{{P}^{g}=\left [ \zeta_{1}^{g}, \zeta_{2}^{g},\cdot \cdot \cdot, \zeta_{K_{A}}^{g}\right ]^{T},}
\end{eqnarray}
where \([\cdot]^{T}\) denotes transposition, \(\zeta_{\varrho}^{g}\in\Upsilon\) for \(\varrho =1, 2,..., K_A\), \({P}^{g}\in\Upsilon^{K_A}\), and \(\Upsilon\) denotes the \(M\)-ary symbol alphabet.  To maintain the average transmission power per SEFDM-IM symbol at unity, \(\Upsilon\) is scaled by \(\sqrt{K/K_A}\). After this point, the \(g\)-th subblock at the output of the symbol mapper is constructed from
\begin{eqnarray}\label{eq:transmitted_message}
{{S}^{g}=\overline{\boldsymbol{\mathbf{I}}}_{K\times K_{A}}^{g}\cdot{P}^{g},
}
\end{eqnarray}
where \(\overline{\mathbf{I}}_{K\times K_{A}}^{g}\) is a \(K\times K_{A}\) activation matrix whose columns are drawn from a \(K\times K\) identity matrix for those indices corresponding to activated subcarriers. For example, the first subblock (i.e., \(g=1\)) with an activation pattern of \(I^{1}=\left \{ 1,0,1,0 \right\}\) is given by  
\begin{equation}\label{eq:tx_message_example}
  \begin{aligned}
    S^{1}&=\begin{bmatrix}
 1&0 \\ 
 0&0 \\ 
 0&1 \\ 
 0&0 
\end{bmatrix}\cdot \begin{bmatrix}
\zeta_{1}^{1}\\ 
\zeta_{2}^{1}
\end{bmatrix}\\
    &=\left [ \zeta_{1}^{1}, 0,\zeta_{2}^{1},0 \right ]^{T}.
  \end{aligned}
\end{equation}

Consequently, a total of \(L=L_1+L_2\) bits are conveyed by one subblock. An SEFDM-IM block is formed by \(G\) concatenated subblocks, expressed as
\begin{equation}\label{eq:concatenation}
  \begin{aligned}
    {S}&={\left [ {S}^{1},{S}^{2},\cdot \cdot \cdot ,{S}^{G} \right ]}^{T}\\
    &={\left [ s_{1}, s_{2},\cdot \cdot \cdot , s_{N} \right ]}^{T},
  \end{aligned}
\end{equation}
where \(s_{k}\in\left \{ 0,\Upsilon \right \}\) for \(k=1,2,...,N\), and \(N=KG\) is the total number of subcarriers. A discrete $N$-subcarrier modulated SEFDM-IM signal is expressed as
\begin{eqnarray}\label{eq:SEFDM_modulation}
{x_{n}=\frac{1}{\sqrt{N}}\sum_{k=1}^{N}s_{k}e^{j2\pi\alpha\frac{kn}{N} },}
\end{eqnarray}
for \(n=1,2,...,N\), where \(1/\sqrt{N}\) is the power normalization factor. \(\alpha =\Delta fT\) is the bandwidth compression factor, where \(\Delta f\) is the subcarrier spacing and \(T\) is the SEFDM-IM symbol duration. \(\alpha<1\) indicates an SEFDM signal, which is converted to a traditional OFDM signal when \(\alpha =1\). SEFDM modulation can be performed by a bank of modulators operating on the non-orthogonal subcarrier frequencies, given by 
\begin{eqnarray}\label{eq:SEFDM_modulation_matrix}
{{X}=\boldsymbol{\Phi}\cdot  {S},}
\end{eqnarray}
where \({X}={\left [ x_{1},x_{2},\cdot \cdot \cdot ,x_{N} \right ]}^{T}\), and \(\boldsymbol{\Phi}\) is an \(N\times N\) carrier matrix whose elements are given by \(\boldsymbol{\Phi}_{k,n}=(1/{\sqrt{N}})e^{j2\pi\alpha\frac{kn}{N} }\).

In practice, channel coding is considered leading to an achievable spectral efficiency of an SEFDM-IM system as
\begin{eqnarray}\label{eq:coded_SE}
{\textnormal{SE}=\frac{1}{\alpha}\frac{\mathcal{R}L}{K}.}
\end{eqnarray}
It is clear that the spectral efficiency is increased due to the bandwidth compression factor $\alpha$.

The PAPR of the SEFDM-IM signal is calculated by
\begin{eqnarray}\label{eq:papr}
{\textnormal{PAPR}=\frac{\max \left \{ \left | x_{n} \right |^{2} \right \}}{E\left \{ \left | x_{n} \right |^{2} \right \}},}
\end{eqnarray}
for \(n=1,2,...,N\), where \(E\left \{ \cdot  \right \}\) denotes the expectation operator. The numerator and the denominator in \eqref{eq:papr} yield the peak and average power of the signal, respectively, and both of them are dependant on the input symbol vector \(S\). Considering its random nature, PAPR is described statistically by the complementary cumulative distribution function (CCDF), given by
\begin{eqnarray}\label{eq:ccdf}
{\textnormal{CCDF}_{\Gamma}\left ( \gamma \right )=\textnormal{Pr}\left ( \Gamma> \gamma \right ),}
\end{eqnarray}
which calculates the probability of the PAPR of a transmitted symbol exceeding a given threshold \(\gamma\), and \(\Pr\left (\cdot\right )\) denotes the probability of an event.

\subsection{Receiver Design}

The signal is transmitted over an AWGN channel, and the received signal is given by 
\begin{eqnarray}\label{eq:received_signal}
{{Y}={X}+{W},}
\end{eqnarray}
where the noise component \({W}={\left [ w_{1},w_{2},\cdot \cdot \cdot ,w_{N} \right ]}^{T}\) comprises \(N\) noise samples drawn from a complex Gaussian distribution \(\mathcal{CN}\left ( 0,N_{0} \right )\) and \(N_{0}\) is the noise variance.
The received signal is projected onto the conjugate of non-orthogonal subcarriers, and the demodulated signal is 
\begin{equation}\label{eq:matched_filtering}
  \begin{aligned}
    {R}&=\boldsymbol{\Phi}^{H}\cdot {Y}\\
    &=\boldsymbol{C}\cdot S+{W}_{\boldsymbol{\Phi}^{H}},
  \end{aligned}
\end{equation}
where \([\cdot]^{H}\) denotes Hermitian transposition, \({W}_{\boldsymbol{\Phi}^{H}}\) corresponds to demodulated noise samples, and \(\boldsymbol{C}\) is the correlation matrix given by \(\boldsymbol{C}=\boldsymbol{\Phi}^{H}\boldsymbol{\Phi}\).

The non-zero off-diagonal elements in the correlation matrix \(\boldsymbol{C}\) characterise the ICI caused by non-orthogonal subcarriers, which are given by
\begin{eqnarray}\label{eq:correlation_matrix}
{\boldsymbol{C}_{k,n}=\begin{cases}
 1& ,k=n \\ 
 \frac{1}{N}\frac{1-e^{j2\pi\alpha(k-n)}}{1-e^{\frac{j2\pi\alpha(k-n)}{N}}}& ,k\neq n 
\end{cases}.}
\end{eqnarray}

After de-modulation, soft information of coded bits is required for LDPC decoding. Due to ICI, the optimal detection needs to consider \(N\) subcarriers, resulting in high computational complexity for large \(N\). The ICI in SEFDM-IM can be divided into two parts: the intra-subblock ICI caused by subcarriers within the same subblock, and the inter-subblock ICI caused by subcarriers in different subblocks. This work proposes efficient solutions to mitigate inter-subblock ICI and enables subblock-based detection. Hence, the received SEFDM-IM symbol is split into \(G\) subblocks in the IM block splitter, and we assume that each demodulated subblock is independent. Next, the LLR calculator gives the natural logarithm of the ratio of probabilities of a 0 being transmitted versus a 1 being transmitted based on the received signal. We define \(\mathcal{I}_{l,0}\) and \(\mathcal{I}_{l,1}\) as the subsets of \(\mathcal{I}\) that transmit a 0 and a 1 as the \(l\)-th index bit for \(l=1,2,...,L_1\), respectively. Likewise, \(\Upsilon_{v,0}^{K_A}\) and \(\Upsilon_{v,1}^{K_A}\) denote the subsets of \(\Upsilon^{K_A}\) that transmit a 0 and a 1 as the \(v\)-th data bit for \(v=1,2,...,L_2\), respectively. 
The \(l\)-th coded index bit in the \(g\)-th subblock is denoted as \({\mathcal{C}_1}^{g}\left ( l \right )\), and its LLR value is given by
\begin{eqnarray}\label{eq:LLR_index_bits}
{\lambda \left ( {\mathcal{C}_1}^{g}\left ( l \right )|{R}^{g} \right )=\ln\frac{\Pr\left ( {\mathcal{C}_1}^{g}\left ( l \right )=0|{R}^{g} \right )}{\Pr\left ( {\mathcal{C}_1}^{g}\left ( l \right )=1|{R}^{g} \right )},}
\end{eqnarray}
where \({R}^{g}\) is the \(g\)-th \(K\times1\) vector of the demodulated signal. Assuming the same a priori probabilities for all valid activation patterns and data symbols, \eqref{eq:LLR_index_bits} can be expressed as 
\begin{eqnarray}\label{eq:LLR_index_2}
{\lambda \left ( {\mathcal{C}_1}^{g}\left ( l \right )|{R}^{g} \right )=\ln\frac{\underset{{{I}}^{g}\in \mathcal{I}_{l,0}}\Sigma\;\underset{{{P}}^{g}\in \Upsilon^{K_A}}\Sigma \Pr\left ( {R}^{g}|{S}^{g} \right )}{\underset{{{I}}^{g}\in \mathcal{I}_{l,1}}\Sigma\;\underset{{{P}}^{g}\in \Upsilon^{K_A}}\Sigma \Pr\left ( {R}^{g}|{S}^{g} \right )},}
\end{eqnarray}
where the transmitted subblock \({S}^{g}\) is constructed from \({I}^{g}\) and \({P}^{g}\) via \eqref{eq:transmitted_message}. The likelihood function for the \(g\)-th subblock is formulated as
\begin{equation}\label{eq:likelihood}
  \begin{aligned}
    \Pr\left ( {R}^{g}|{S}^{g} \right )&=\frac{e^{-\frac{1}{N_{o}}\left ( {R}^{g}-\boldsymbol{C}^{g}{S}^{g} \right )^{H}\left ( {R}^{g}-\boldsymbol{C}^{g}{S}^{g} \right )}}{\pi N_{0}}\\
    &=\frac{e^{-\Psi\left ( {I}^{g},{P}^{g} \right )}}{\pi N_{0}},
  \end{aligned}
\end{equation}
where \(\boldsymbol{C}^{g}\) is the \(g\)-th \(K\times K\) sub-matrix of the correlation matrix. For brevity of presentation, \(\Psi\left ( {I}^{g},{P}^{g} \right )\) is defined as
\begin{eqnarray}\label{eq:Psi_white}
{\Psi\left ( {I}^{g},{P}^{g} \right )=\frac{1}{N_0}\left ( {R}^{g}-\boldsymbol{C}^{g}{S}^{g} \right )^{H}\left ( {R}^{g}-\boldsymbol{C}^{g}{S}^{g} \right ).}
\end{eqnarray}

Considering \eqref{eq:likelihood} and \eqref{eq:Psi_white}, the expression in \eqref{eq:LLR_index_2} is simplified to
\begin{eqnarray}\label{eq:LLR_index_3}
{\lambda \left ({\mathcal{C}_1}^{g}\left ( l \right )|{R}^{g} \right )=\ln\frac{\underset{{{I}}^{g}\in \mathcal{I}_{l,0}}\Sigma\;\underset{{{P}}^{g}\in \Upsilon^{K_A}}\Sigma e^{ -{\Psi}\left ( {I}^{g},{P}^{g} \right ) }}{\underset{{{I}}^{g}\in \mathcal{I}_{l,1}}\Sigma\;\underset{{{P}}^{g}\in \Upsilon^{K_A}}\Sigma e^{ -{\Psi}\left ( {I}^{g},{P}^{g} \right ) }}. }
\end{eqnarray}

Since the indices of activated subcarriers are not known at the receiver, all valid activation patterns need to be considered when calculating soft information for coded data bits. The \(v\)-th coded data bit in the \(g\)-th subblock is denoted as \({\mathcal{C}_2}^{g}\left ( v \right )\), and its LLR value is given by
\begin{eqnarray}\label{eq:LLR_data_1}
{\lambda \left ({\mathcal{C}_2}^{g}\left ( v \right )|{R}^{g} \right )=\ln\frac{\Pr\left ({\mathcal{C}_2}^{g}\left ( v \right )=0|{R}^{g} \right )}{\Pr\left ({\mathcal{C}_2}^{g}\left ( v \right )=1|{R}^{g} \right )}.}
\end{eqnarray}

Then, similar to the LLR calculations for coded index bits, \eqref{eq:LLR_data_1} simplifies to
\begin{equation}\label{eq:LLR_data_2}
  \begin{aligned}
    \lambda \left ( {\mathcal{C}_2}^{g}\left ( v \right )|{R}^{g} \right )&=\ln\frac{\underset{{{I}}^{g}\in \mathcal{I}}\Sigma\;\underset{{{P}}^{g}\in \Upsilon_{v,0}^{K_A}}\Sigma \Pr\left ( {R}^{g}|{S}^{g} \right )}{\underset{{{I}}^{g}\in \mathcal{I}}\Sigma\;\underset{{{P}}^{g}\in \Upsilon_{v,1}^{K_A}}\Sigma \Pr\left ( {R}^{g}|{S}^{g} \right )}\\
    &=\ln\frac{\underset{{{I}}^{g}\in \mathcal{I}}\Sigma\;\underset{{{P}}^{g}\in \Upsilon_{v,0}^{K_A}}\Sigma e^{ -{\Psi}\left ( {I}^{g},{P}^{g} \right ) }}{\underset{{{I}}^{g}\in \mathcal{I}}\Sigma\;\underset{{{P}}^{g}\in \Upsilon_{v,1}^{K_A}}\Sigma e^{ -{\Psi}\left ( {I}^{g},{P}^{g} \right ) }}. 
  \end{aligned}
\end{equation}

A low complexity LLR calculation method could be used according to \cite{low_complexity_llr_calculation}. This work mainly considers uplink NB-IoT applications, therefore optimal LLR calculation is acceptable as most of complicated signal processing is within the central point, which is not sensitive to signal processing complexity.
After collecting the LLR values for coded bits in each subblock, two LDPC decoders perform decoding for index bit sequence and data bit sequence separately. Then the two sequences of decoded bits \(\widehat{{{\mathcal{B}}_1}}\) and \(\widehat{{{\mathcal{B}}_2}}\) are formulated into one output bit sequence \(\widehat{{{\mathcal{B}}}}\) in the bit combiner.

\section{Proposed Subcarrier Pattern Designs}\label{section2}

We define subcarrier patterns as the combination of activation patterns and modulation schemes. In this section, we present both traditional and proposed subcarrier pattern designs for SEFDM-IM and demonstrate the superiority of our proposals.

\subsection{Challenges for Existing IM Systems}

The traditional pattern design principle deployed in existing work \cite{SEFDMIM_CHINA,SEFDMIM_JAPAN,SEFDM-IM_jointChannel} have three characteristics: first, there is a fixed number of activated subcarriers in one subblock, and second, all subcarriers in one subblock are probable to be activated. Lastly, all activated subcarriers are modulated with the same modulation scheme. In other words, both \(K_A\) and \(M\) have a fixed value. An example lookup table for traditional subcarrier patterns in SEFDM-IM with \([K,K_A]=[4,1]\) is given in Table \ref{tab:41_traditional_SEFDMIM}, where the values of \(K\) and \(K_A\) are specified in brackets for notational convenience. In four valid patterns, one out of four subcarriers is activated and modulated with a data symbol \(\mathcal{S}_A^{\left ( 1 \right )}\), where the subscript \(A\) denotes the modulation cardinality \(M_A\) (e.g., \(M_A=4\) for QPSK) and the superscript in brackets corresponds to the indices of symbols in the data symbol vector \({{P}}^{g}\). In this case, \({{P}}^{g}\) has only one symbol for \(K_A=1\). According to \eqref{eq:correlation_matrix}, the ICI level increases as the frequency spacing between two activated subcarriers decreases. Severe inter-subblock ICI is introduced when the last subcarrier in one subblock and the first subcarrier in the following subblock are both activated. In this case, subblock-based detection that ignores inter-subblock ICI for computational efficiency suffers from performance degradation. For SEFDM-IM systems with more activated subcarriers, i.e., lower values of \(K/K_A\), the probability of having adjacently-located activated subcarriers in neighbouring subblocks is increased, leading to a higher level of inter-subblock ICI impairment.

\subsection{Proposed Designs in this Work}

\begin{table}[]
\centering
\caption{\\Subcarrier pattern lookup table for traditional SEFDM-IM with \([K,K_A ]= [4,1 ]\).}
\begin{tabular}{|c|c|c|c|}
\hline&&&\\[-0.65em]
\textit{\textbf{Pattern}} & \textit{\textbf{Index bits}} & \textit{\textbf{Activation patterns}} & \textit{\textbf{Subcarrier patterns}} \\ [0.5ex] 
\hline\hline &&&\\[-0.65em]
1  & {[}0,\;0{]}  & \{1,\;0,\;0,\;0\} & \(\left [ \mathcal{S}_A^{\left ( 1 \right )},\;0,\;0,\;0 \right ]^{T}\) \\ [1.1ex] \hline&&&\\[-0.65em]
2  & {[}0,\;1{]}  & \{0,\;0,\;0,\;1\} & \(\left [ 0,\;0,\;0,\;\mathcal{S}_A^{\left ( 1 \right )} \right ]^{T}\) \\ [1.1ex]  \hline&&&\\[-0.65em]
3  & {[}1,\;0{]} & \{0,\;1,\;0,\;0\} & \(\left [ 0,\;\mathcal{S}_A^{\left ( 1 \right )},\;0,\;0 \right ]^{T}\) \\ [1.1ex] \hline&&&\\[-0.65em]
4  & {[}1,\;1{]}  & \{0,\;0,\;1,\;0\} & \(\left [ 0,\;0,\;\mathcal{S}_A^{\left ( 1 \right )},\;0 \right ]^{T}\) \\ [1.1ex] 
\hline
\end{tabular}
\label{tab:41_traditional_SEFDMIM}
\end{table}

Against the above background, we propose a novel pattern design principle for SEFDM-IM, where the last subcarrier in each subblock is always left unused. As a result, activated subcarriers in neighbouring subblocks are separated by at least one unused subcarrier, leading to reduced inter-subblock ICI. A similar method, termed multiband SEFDM, was shown to be efficacious in improving and simplifying SEFDM detection in \cite{TongyangCSNDSPBSEFDM}. An alternative approach to mitigate inter-subblock interference is to optimize spectrum features via cutting out-of-band power emissions, which was explored in \cite{psd}. The method proposed in this work is more flexible since a partial number of subcarriers are activated leading to unique activation patterns in Table \ref{tab:412_SEFDMIM}, where pattern-1, pattern-3 and pattern-4 have the same subcarrier patterns as those of the traditional SEFDM-IM with \([K,K_A]=[4,1]\). Advantageously, pattern-2 simultaneously activates two subcarriers, the first and the third, whilst keeping the last subcarrier space void of energy,  i.e., \(K_A=2\). Consequently, index bits are transmitted by both the number of activated subcarriers and their locations. To avoid error propagation due to incorrect detection of activation patterns, \(L\) is fixed regardless of the \(K_A\) value. More specifically, the number of data bits transmitted in pattern-2 should be the same as that of other patterns. The extra activated subcarrier provides a new degree of freedom to alter the subcarrier pattern in pattern-2. For research completeness and convincing comparisons, three possible realizations of subcarrier pattern designs are presented as follows. It should be noted that \(M_A\) is the modulation cardinality used in pattern-1, pattern-3 and pattern-4.
\begin{itemize}
    \item {\emph{Proposed SEFDM-IM-1}: In pattern-2, the first subcarrier is modulated with a pre-defined signaling symbol \( \mathcal{S}_A^{\left ( * \right )}\) known at the receiver, and the third subcarrier is modulated with a data symbol of the same modulation cardinality \(M_A\) mapped from data bits.}
    \item{\emph{Proposed SEFDM-IM-2}: Motivated by repetition coding, in pattern-2, the first and the third subcarriers are modulated with the same data symbol mapped from data bits.}
    \item{\emph{Proposed SEFDM-IM-3}: In pattern-2, the first and the third subcarriers are modulated with two data symbols \( \mathcal{S}_B^{\left ( 1 \right )}\) and \( \mathcal{S}_C^{\left ( 1 \right )}\) of the modulation cardinalities \(M_B\) and \(M_C\), which satisfies the condition
    \begin{eqnarray}\label{eq:412_condition1}
    {M_{B}M_{C}=M_{A}},
    \end{eqnarray}
    which ensures that \(L_2\) data bits are conveyed by pattern-2. In addition, to help the receiver distinguish between pattern-2 and the other patterns, \(M_A\), \(M_B\) and \(M_C\) satisfy
    \begin{eqnarray}\label{eq:412_condition2}
    {\left ( M_{B}\neq M_{A} \right )\lor\left ( M_{C}\neq M_{A} \right )= 1},
    \end{eqnarray}    
    where \(\lor\) stands for the logical OR operator. \eqref{eq:412_condition2} indicates that there is at least one modulation cardinality used in pattern-2 that is different from \(M_A\).}
\end{itemize}
When the traditional subcarrier pattern design is deployed, \(M\) and \(K_A\) have a fixed value regardless of the selected activation pattern, and hence data symbol vector \({{P}}^{g}\) is independent from activation patterns. By contrast, for the proposed subcarrier pattern design, index bits determine the activation pattern for each subblock, which in turn determines the values of \(M\) and \(K_A\) for constructing \({{P}}^{g}\).

\begin{table}[]
\centering
\caption{\\Subcarrier pattern lookup table for proposed SEFDM-IM with \([K,K_A]=[4,(1,2)]\). }
\begin{tabular*}{\columnwidth}{ @{\extracolsep{\fill}} |c|c|c|c|}
\hline&&&\\[-0.65em]
\textit{\textbf{Pattern}} & \textit{\textbf{\begin{tabular}[c]{@{}c@{}}Index\\ bits\end{tabular}}} & \textit{\textbf{\begin{tabular}[c]{@{}c@{}}Activation\\ patterns\end{tabular}}} & \textit{\textbf{Subcarrier patterns}} \\ [0.5ex] 
\hline\hline &&&\\[-0.65em]
1  & {[}0,\;0{]}  & \{1,\;0,\;0,\;0\} & \(\left [ \mathcal{S}_A^{\left ( 1 \right )},\;0,\;0,\;0 \right ]^{T}\) \\ [1.1ex] \hline&&&\\[-0.65em]
2  & {[}0,\;1{]}  & \{1,\;0,\;1,\;0\} & 
\begin{tabular}[c]{@{}c@{}}
SEFDM-IM-1:\(\left [ \mathcal{S}_A^{\left ( * \right )},\;0,\;\mathcal{S}_A^{\left ( 1 \right )},\;0 \right ]^{T}\)\\ [0.8ex]
SEFDM-IM-2:\(\left [ \mathcal{S}_A^{\left ( 1 \right )},\;0,\;\mathcal{S}_A^{\left ( 1 \right )},\;0 \right ]^{T}\)\\ [0.8ex]
SEFDM-IM-3:\(\left [ \mathcal{S}_B^{\left ( 1 \right )},\;0,\;\mathcal{S}_C^{\left ( 1 \right )},\;0 \right ]^{T}\)\\ [1ex]
\end{tabular}
\\ [1.1ex]  
\hline&&&\\[-0.65em]
3  & {[}1,\;0{]} & \{0,\;1,\;0,\;0\} & \(\left [ 0,\;\mathcal{S}_A^{\left ( 1 \right )},\;0,\;0 \right ]^{T}\) \\ [1.1ex] \hline&&&\\[-0.65em]
4  & {[}1,\;1{]}  & \{0,\;0,\;1,\;0\} & \(\left [ 0,\;0,\;\mathcal{S}_A^{\left ( 1 \right )},\;0 \right ]^{T}\) \\ [1.1ex] 
\hline
\end{tabular*}
\label{tab:412_SEFDMIM}
\end{table}

We also propose subcarrier patterns for a larger number of activated subcarriers, which is shown in Table \ref{tab:423_SEFDMIM}. Explicitly, two out of four subcarriers are activated and modulated with two \(M_A\)-ary data symbols for pattern-1, pattern-3 and pattern-4. Similarly, we propose three subcarrier patterns for pattern-2, where three subcarriers are activated simultaneously. The extra activated subcarrier is either modulated with a pre-defined signaling symbol \(\mathcal{S}_A^{\left ( * \right )}\) for SEFDM-IM-1 or the repetition of the first data symbol \(\mathcal{S}_A^{\left ( 1 \right )}\) for SEFDM-IM-2. For SEFDM-IM-3, three modulation cardinalities \(M_{B}\), \(M_{C}\) and \(M_{D}\) are deployed, and they satisfy two following conditions
\begin{eqnarray}\label{eq:condition1}
{M_{B}M_{C}M_{D}=\left (  M_{A}\right )^{2}},
\end{eqnarray}
and
\begin{eqnarray}\label{eq:condition2}
{\left ( M_{B}\neq M_{A} \right )\lor\left ( M_{C}\neq M_{A} \right )\lor\left ( M_{D}\neq M_{A} \right )= 1}.
\end{eqnarray}
These conditions ensure a constant number of bits transmitted in pattern-2 and at least one distinct modulation cardinality used in pattern-2.

The novel subcarrier pattern designs provided above are proposed from the point of view of activation patterns, which also could be used in other non-orthogonal systems such as multi-carrier faster-than-Nyquist (MFTN) \cite{MFTN}. Given that modulation schemes have an impact on index-modulated system performance, we also consider new pattern designs from the perspective of modulation schemes. While existing work only considers BPSK and QPSK modulation, we explore higher modulation cardinalities up to 16QAM, which can be combined with both traditional and proposed pattern design principles.

\begin{table}[]
\centering
\caption{\\Subcarrier pattern lookup table for proposed SEFDM-IM with \([K,K_A]=[4,(2,3)]\).}
\begin{tabular}{|c|c|c|c|}
\hline&&&\\[-0.65em]
\textit{\textbf{Pattern}} & \textit{\textbf{\begin{tabular}[c]{@{}c@{}}Index\\ bits\end{tabular}}} & \textit{\textbf{\begin{tabular}[c]{@{}c@{}}Activation\\ patterns\end{tabular}}} & \textit{\textbf{Subcarrier patterns}} \\ [0.5ex] 
\hline\hline &&&\\[-0.65em]
1  & {[}0,\;0{]}  & \{0,\;1,\;1,\;0\} & \(\left [0,\; \mathcal{S}_A^{\left ( 1 \right )},\;\mathcal{S}_A^{\left ( 2 \right )},\;0 \right ]^{T}\) \\ [1.1ex] \hline&&&\\[-0.65em]
2  & {[}0,\;1{]}  & \{1,\;1,\;1,\;0\} & 
\begin{tabular}[c]{@{}c@{}}
SEFDM-IM-1:\(\left [ \mathcal{S}_A^{\left ( * \right )},\mathcal{S}_A^{\left ( 1 \right )},\mathcal{S}_A^{\left ( 2 \right )},0 \right ]^{T}\)\\ [0.8ex]
SEFDM-IM-2:\(\left [ \mathcal{S}_A^{\left ( 1 \right )},\mathcal{S}_A^{\left ( 1 \right )},\mathcal{S}_A^{\left ( 2 \right )},0 \right ]^{T}\)\\ [0.8ex]
SEFDM-IM-3:\(\left [ \mathcal{S}_B^{\left ( 1 \right )},\mathcal{S}_C^{\left ( 1 \right )},\mathcal{S}_D^{\left ( 1 \right )},0 \right ]^{T}\)\\ [1ex]
\end{tabular}
\\ [0.5ex]  
\hline&&&\\[-0.65em]
3  & {[}1,\;0{]} & \{1,\;0,\;1,\;0\} & \(\left [ \mathcal{S}_A^{\left ( 1 \right )},\;0,\;\mathcal{S}_A^{\left ( 2 \right )},\;0 \right ]^{T}\) \\ [1.1ex] \hline&&&\\[-0.65em]
4  & {[}1,\;1{]}  & \{1,\;1,\;0,\;0\} & \(\left [ \mathcal{S}_A^{\left ( 1 \right )},\;\mathcal{S}_A^{\left ( 2 \right )},\;0,\;0 \right ]^{T}\) \\ [1.1ex] 
\hline
\end{tabular}
\label{tab:423_SEFDMIM}
\end{table}

\section{Pattern Designs for Selected Spectral Efficiency}\label{section3}

In this section, we propose the subcarrier pattern designs for four selected candidates of spectral efficiency, namely, \(\textnormal{SE}=1.5,\;2,\;2.2,\;2.5\) bit/s/Hz. Noting that these spectral efficiency values do not consider coding rate, i.e., \(\mathcal{R}=1\). 
The spectral efficiency is related to the number of activated subcarriers and the order of modulation schemes. Since the change of these two factors is discrete, the spectral efficiency values we investigate are also discrete instead of being continuous. In each case, different activation patterns and modulation schemes are jointly considered. System configurations are specified by the activation parameters \([K,K_A]\), modulation schemes and the values of \(\alpha\), followed by a detailed description of each subcarrier pattern. For brevity, traditional SEFDM-IM schemes that have been deployed in previous research are denoted as "SEFDM-IM-Tra", where "Tra" is the abbreviation for "Traditional". In this paper, we only consider \(K=4\), which yields \(U=4\) and \(L_1=2\). The proposed subcarrier pattern designs can be extended to higher values of \(K\) as long as the last subcarrier in each subblock is unused. The maximum modulation format we investigate is 16QAM. For IM systems with even-higher-order modulation schemes, chasing for index-domain via intentionally deleting subcarriers will cause significant spectral efficiency loss. Moreover, this work aims for NB-IoT, where the current 3GPP standard defines the maximum modulation format as QPSK. In the future 3GPP standard, the modulation format could be improved to 16QAM.

\subsection{Spectral efficiency of 1.5 bit/s/Hz}
\begin{itemize}
    \item{ SEFDM-IM-Tra, \([K,K_A]=[4,1]\), QPSK and \(\alpha=0.67\): One out of four subcarriers is activated and modulated with a QPSK data symbol, i.e., \(M_A=4\).}
    \item{ SEFDM-IM-1, \([K,K_A]=[4,(1,2)]\), QPSK and \(\alpha=0.67\): In pattern-1, pattern-3 and pattern-4, the single activated subcarrier is modulated with a QPSK data symbol. In pattern-2, one subcarrier is modulated with a QPSK data symbol and the other is modulated with a pre-defined QPSK signalling symbol known at the receiver.}
    \item{ SEFDM-IM-2, \([K,K_A]=[4,(1,2)]\), QPSK and \(\alpha=0.67\): In pattern-1, pattern-3 and pattern-4, the single activated subcarrier is modulated with a QPSK data symbol. In pattern-2, two subcarriers are modulated with the same QPSK data symbol.}
    \item{ SEFDM-IM-3, \([K,K_A]=[4,(1,2)]\), QPSK and \(\alpha=0.67\): In pattern-1, pattern-3 and pattern-4, the single activated subcarrier is modulated with a QPSK data symbol. In pattern-2, two subcarriers are modulated with two BPSK data symbols, i.e., \(M_B=M_C=2\).}  
\end{itemize}

\subsection{Spectral efficiency of 2 bit/s/Hz}
\begin{itemize}
    \item{ SEFDM-IM-M1, \([K,K_A]=[4,1]\), 8QAM and \(\alpha=0.625\): This is the proposed scheme from a modulation prospective. One out of four subcarriers is activated and modulated with an 8QAM data symbol, i.e., \(M_A=8\).}
    \item{ SEFDM-IM-1, \([K,K_A]=[4,(1,2)]\), 8QAM and \(\alpha=0.625\): In pattern-1, pattern-3 and pattern-4, the single activated subcarrier is modulated with an 8QAM data symbol. In pattern-2, one subcarrier is modulated with an 8QAM data symbol and the other is modulated with a pre-defined 8QAM signalling symbol known at the receiver.}
    \item{ SEFDM-IM-2, \([K,K_A]=[4,(1,2)]\), 8QAM and \(\alpha=0.625\): In pattern-1, pattern-3 and pattern-4, the single activated subcarrier is modulated with an 8QAM data symbol. In pattern-2, two subcarriers are modulated with the same 8QAM data symbol.}
    \item{ SEFDM-IM-3, \([K,K_A]=[4,(1,2)]\), 8QAM and \(\alpha=0.625\): In pattern-1, pattern-3 and pattern-4, the single activated subcarrier is modulated with an 8QAM data symbol. In pattern-2, two subcarriers are modulated with a QPSK and a BPSK data symbols, i.e., \(M_B=4\) and \(M_C=2\).}  
    \item{ SEFDM-IM-Tra, \([K,K_A]=[4,2]\), QPSK and \(\alpha=0.75\): Two out of four subcarriers are activated and modulated with two QPSK data symbols, i.e., \(M_A=4\).}    
    \item{ SEFDM-IM-1, \([K,K_A]=[4,(2,3)]\), QPSK and \(\alpha=0.75\): In pattern-1, pattern-3 and pattern-4, two activated subcarriers are modulated with two QPSK data symbols. In pattern-2, the second and the third subcarriers are modulated with QPSK data symbols, while the first subcarrier is modulated with a pre-defined QPSK signalling symbol.}
    \item{ SEFDM-IM-2, \([K,K_A]=[4,(2,3)]\), QPSK and \(\alpha=0.75\): In pattern-1, pattern-3 and pattern-4, two activated subcarriers are modulated with two QPSK data symbols. In pattern-2, the second and the third subcarriers are modulated with QPSK data symbols, while the first subcarrier is modulated with the same data symbol as the second subcarrier.}
    \item{ SEFDM-IM-3, \([K,K_A]=[4,(2,3)]\), QPSK and \(\alpha=0.75\): In pattern-1, pattern-3 and pattern-4, two activated subcarriers are modulated with two QPSK data symbols. In pattern-2, the first and the third subcarriers are modulated with BPSK data symbols, i.e., \(M_B=M_D=2\), while the second subcarrier is modulated a QPSK data symbol, i.e., \(M_C=4\).}
\end{itemize}

\begin{table}[]
\centering
\caption{\\Subcarrier pattern lookup table for traditional SEFDM-IM with \([K,K_A]=[4,3]\).}
\begin{tabular}{|c|c|c|c|}
\hline&&&\\[-0.65em]
\textit{\textbf{Pattern}} & \textit{\textbf{\begin{tabular}[c]{@{}c@{}}Index\\ bits\end{tabular}}} & \textit{\textbf{Activation patterns}}  & \textit{\textbf{Subcarrier patterns}} \\ [0.5ex] 
\hline\hline &&&\\[-0.65em]
1  & {[}0,\;0{]}  & \{0,\;1,\;1,\;1\} & \(\left [ 0,\;\mathcal{S}_A^{\left ( 1 \right )},\;\mathcal{S}_A^{\left ( 2 \right )},\;\mathcal{S}_A^{\left ( 3 \right )} \right ]^{T}\) \\ [1.1ex] \hline&&&\\[-0.65em]
2  & {[}0,\;1{]}  & \{1,\;1,\;1,\;0\} & 
\(\left [ \mathcal{S}_A^{\left ( 1 \right )},\;\mathcal{S}_A^{\left ( 2 \right )},\;\mathcal{S}_A^{\left ( 3 \right )},\;0 \right ]^{T}\) 
\\ [1.1ex]  
\hline&&&\\[-0.65em]
3  & {[}1,\;0{]} & \{1,\;0,\;1,\;1\} &\(\left [ \mathcal{S}_A^{\left ( 1 \right )},\;0,\;\mathcal{S}_A^{\left ( 2 \right )},\;\mathcal{S}_A^{\left ( 3 \right )} \right ]^{T}\) \\ [1.1ex] \hline&&&\\[-0.65em]
4  & {[}1,\;1{]} & \{1,\;1,\;0,\;1\} & \(\left [ \mathcal{S}_A^{\left ( 1 \right )},\;\mathcal{S}_A^{\left ( 2 \right )},\;0,\;\mathcal{S}_A^{\left ( 3 \right )} \right ]^{T}\) \\ [1.1ex] 
\hline
\end{tabular}
\label{tab:43_SEFDMIM}
\end{table}

\subsection{Spectral efficiency of 2.2 bit/s/Hz}
\begin{itemize}
    \item{ SEFDM-IM-Tra, \([K,K_A]=[4,3]\), QPSK and \(\alpha=0.9\): Subcarrier patterns are given in Table \ref{tab:43_SEFDMIM}. Three out of four subcarriers are activated and modulated with three QPSK data symbols, i.e., \(M_A=4\).}
    \item{ SEFDM-IM-M2, \([K,K_A]=[4,1]\), 16QAM and \(\alpha=0.675\): One out of four subcarriers is activated and modulated with a 16QAM data symbol, i.e., \(M_A=16\).}
    \item{ SEFDM-IM-1, \([K,K_A]=[4,(1,2)]\), 16QAM and \(\alpha=0.675\): In pattern-1, pattern-3 and pattern-4, the single activated subcarrier is modulated with a 16QAM data symbol. In pattern-2, one subcarrier is modulated with a 16QAM data symbol and the other is modulated with a pre-defined 16QAM signalling symbol known at the receiver.}
    \item{ SEFDM-IM-2, \([K,K_A]=[4,(1,2)]\), 16QAM and \(\alpha=0.675\): In pattern-1, pattern-3 and pattern-4, the single activated subcarrier is modulated with a 16QAM data symbol. In pattern-2, two subcarriers are modulated with the same 16QAM data symbol.}
    \item{ SEFDM-IM-3, \([K,K_A]=[4,(1,2)]\), 16QAM and \(\alpha=0.675\): In pattern-1, pattern-3 and pattern-4, the single activated subcarrier is modulated with a 16QAM data symbol. In pattern-2, two subcarriers are modulated with two QPSK data symbols, i.e., \(M_B=M_C=4\).}  
\end{itemize}

\subsection{Spectral efficiency of 2.5 bit/s/Hz}
\begin{itemize}
    \item{ SEFDM-IM-Tra, \([K,K_A]=[4,3]\), QPSK and \(\alpha=0.8\): Three out of four subcarriers are activated and modulated with three QPSK data symbols, i.e., \(M_A=4\).}
    \item{ SEFDM-IM-M2, \([K,K_A]=[4,1]\), 16QAM and \(\alpha=0.6\): One out of four subcarriers is activated and modulated with a 16QAM data symbol, i.e., \(M_A=16\).}
    \item{ SEFDM-IM-1, \([K,K_A]=[4,(1,2)]\), 16QAM and \(\alpha=0.6\): In pattern-1, pattern-3 and pattern-4, the single activated subcarrier is modulated with a 16QAM data symbol. In pattern-2, one subcarrier is modulated with a 16QAM data symbol and the other is modulated with a pre-defined 16QAM signalling symbol known at the receiver.}
    \item{ SEFDM-IM-2, \([K,K_A]=[4,(1,2)]\), 16QAM and \(\alpha=0.6\): In pattern-1, pattern-3 and pattern-4, the single activated subcarrier is modulated with a 16QAM data symbol. In pattern-2, two subcarriers are modulated with the same 16QAM data symbol.}
    \item{ SEFDM-IM-3, \([K,K_A]=[4,(1,2)]\), 16QAM and \(\alpha=0.6\): In pattern-1, pattern-3 and pattern-4, the single activated subcarrier is modulated with a 16QAM data symbol. In pattern-2, two subcarriers are modulated with two QPSK data symbols, i.e., \(M_B=M_C=4\).}  
\end{itemize}

\section{Simulation Results and Discussions}\label{section4}

In this section, we comprehensively investigate BER and PAPR performance for the proposed SEFDM-IM systems. All investigated SEFDM-IM systems have been detailedly described in Section \ref{section3}. The results of classical OFDM-IM and traditional SEFDM-IM-Tra systems are provided as benchmarks.
The performance comparisons of classical OFDM-IM and traditional SEFDM-IM-Tra with basic OFDM have been comprehensively investigated in existing research such as \cite{OFDM_IM_Basar} and \cite{SEFDMIM_JAPAN}. This work aims to improve pattern design in the IM domain. Therefore, we focus our performance comparisons in terms of IM-based schemes.
For convenience, the figure legend is specified by \([K,K_A,\textnormal{modulation scheme}, \alpha]\), where the modulation scheme refers to that used in pattern-1, pattern-3 and pattern-4, i.e., \(M_A\). The modulation schemes used in pattern-2 are not specified, which can be found in Section \ref{section3}. Since we deploy two independent LDPC decoders, we can calculate BER from index bits and data bits separately. By counting the number of differences between the input index bits \({{\mathcal{B}}_1}\) and the output index bits \(\widehat{{{\mathcal{B}}_1}}\), the index BER is obtained. Similarly, the data BER is obtained by comparing \({{\mathcal{B}}_2}\) and \(\widehat{{{\mathcal{B}}_2}}\). The average BER is obtained by comparing \({{\mathcal{B}}}\) and \(\widehat{{{\mathcal{B}}}}\), and it is referred to as BER unless otherwise specified. In all simulations, a coding rate of \(\mathcal{R}=1/2\) is used for LDPC encoding, and both LDPC decoders deploy the belief propagation algorithm with 50 decoding iterations. In addition, we assume an AWGN channel and the following system parameters: \(N=12\) and \(K=4\).

\begin{figure}[]
\begin{center}
\includegraphics[width=\columnwidth]{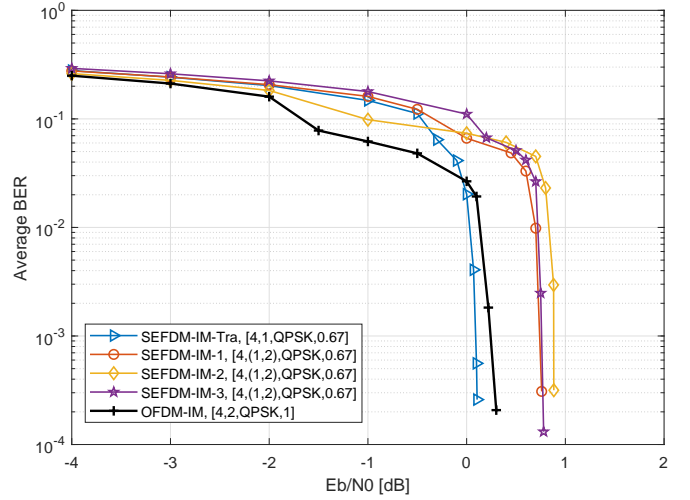}
\end{center}
\caption{Error performance of coded SEFDM-IM and OFDM-IM systems with the spectral efficiency of 0.75 bit/s/Hz.}
\label{Fig:coded_SE15}
\end{figure}

\begin{figure}[]
\begin{center}
\includegraphics[width=\columnwidth]{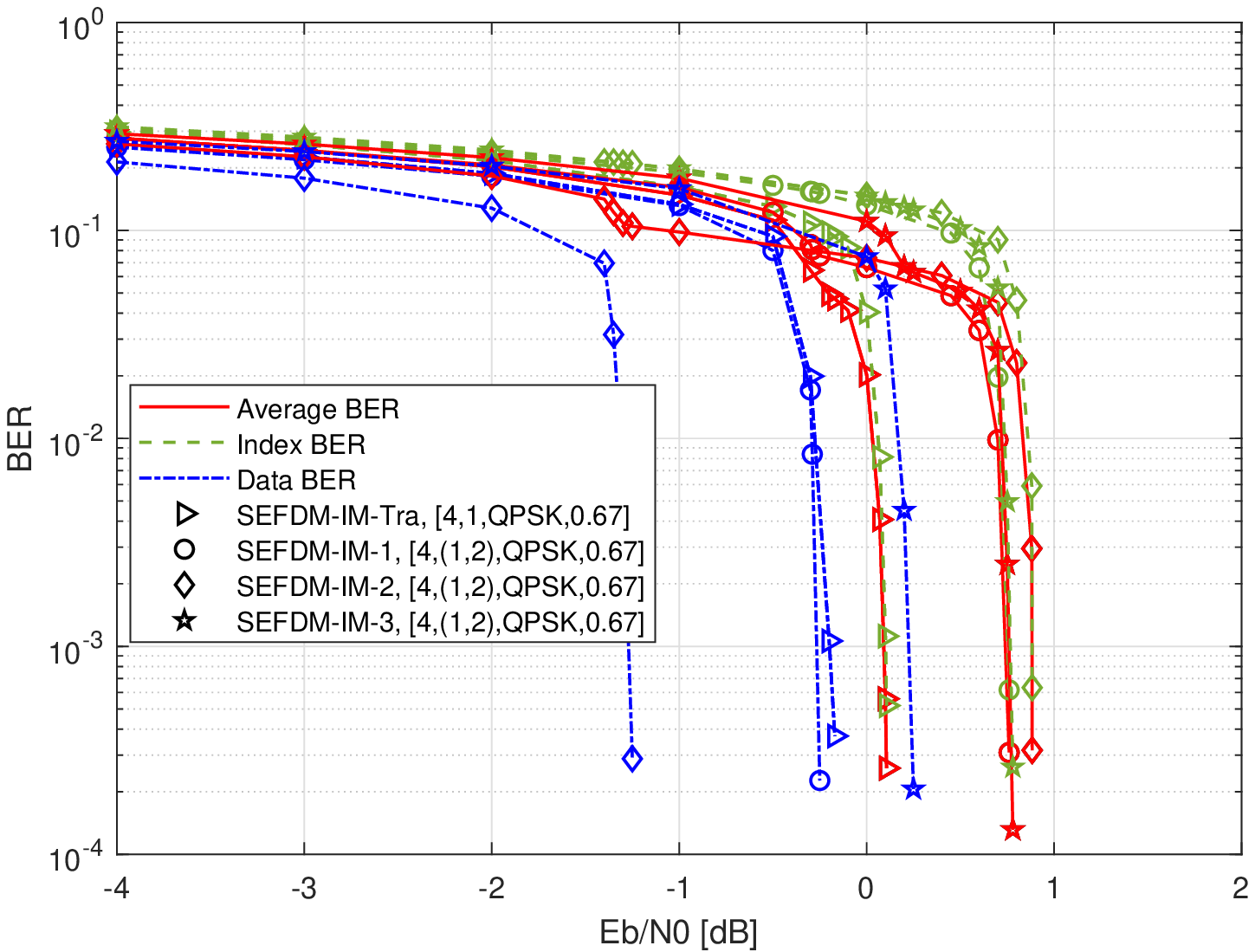}
\end{center}
\caption{Error performance of coded SEFDM-IM systems with the spectral efficiency of 0.75 bit/s/Hz in terms of index BER, data BER and average BER.}
\label{Fig:sep_ber_coded_SE15_all_SEFDMIM}
\end{figure}

\begin{figure}[]
\begin{center}
\includegraphics[width=\columnwidth]{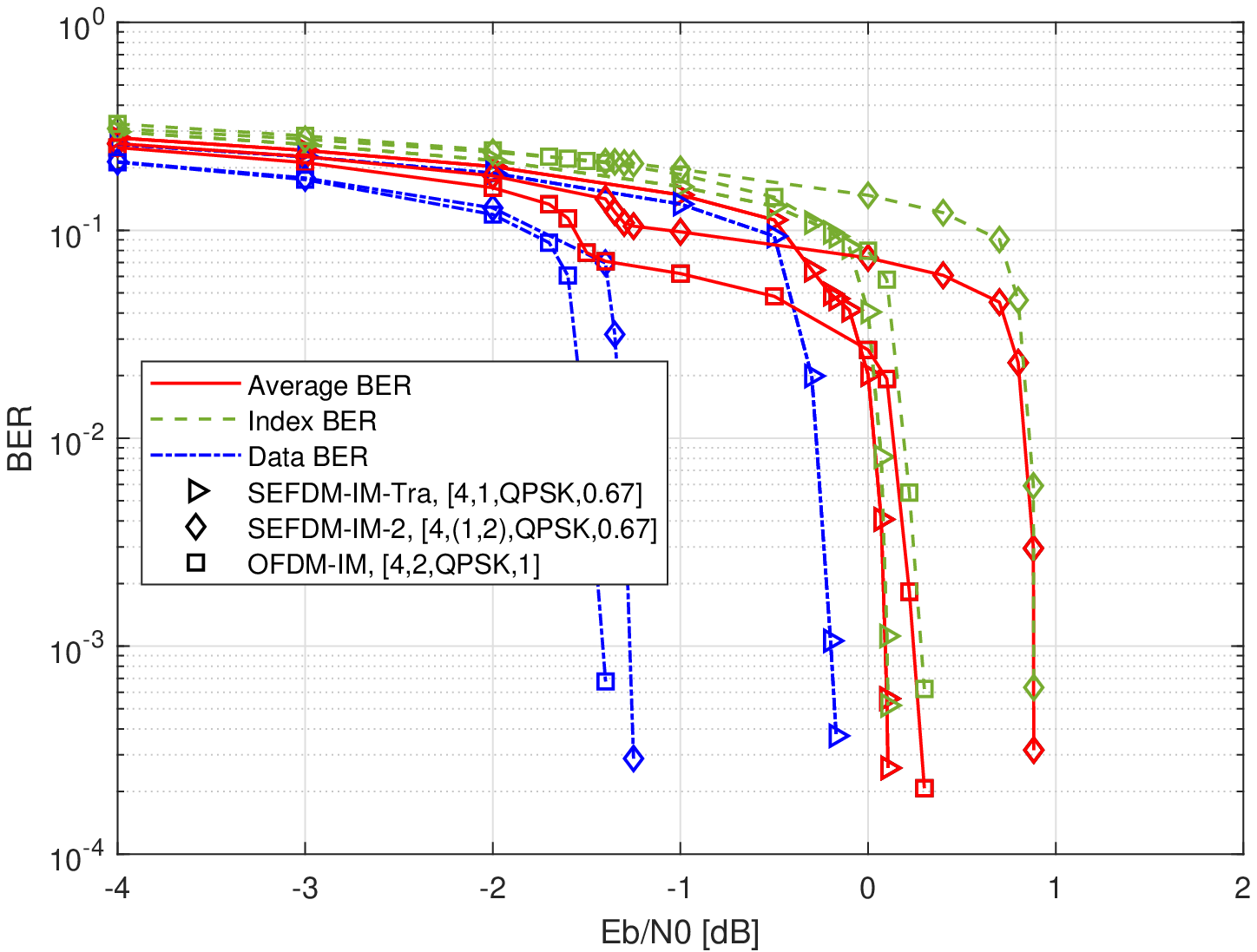}
\end{center}
\caption{Error performance of coded SEFDM-IM and OFDM-IM systems with the spectral efficiency of 0.75 bit/s/Hz in terms of index BER, data BER and average BER.}
\label{Fig:sep_ber_coded_SE15}
\end{figure}

\begin{figure}[]
\begin{center}
\includegraphics[width=\columnwidth]{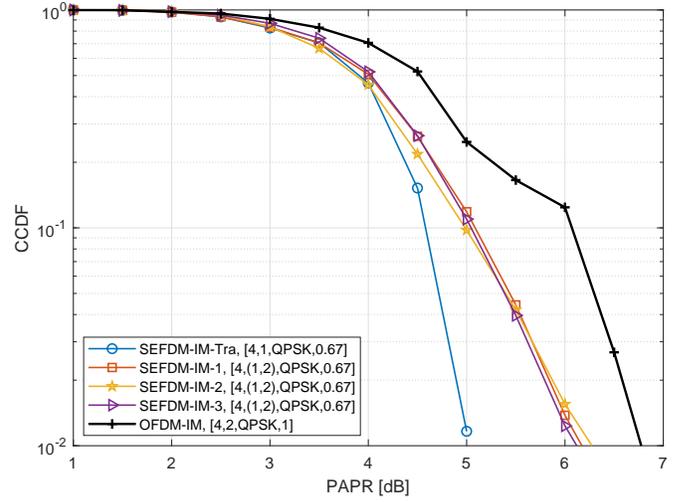}
\end{center}
\caption{CCDF of the PAPR of SEFDM-IM and OFDM-IM systems with the spectral efficiency of 0.75 bit/s/Hz.}
\label{Fig:papr_SE15}
\end{figure}

We first consider SEFDM-IM systems with the spectral efficiency of 1.5 bit/s/Hz, which turns into 0.75 bit/s/Hz after LDPC coding is applied. For fair comparisons, a QPSK modulated OFDM-IM system with \([K,K_A]=[4,2]\) is considered. In Fig. \ref{Fig:coded_SE15}, the BER performance of investigated systems is plotted as a function of \(E_{b}/N_{0}\), and the traditional SEFDM-IM-Tra system with \([K,K_A]=[4,1]\) achieves the best BER performance among all coded systems. As shown in Fig. \ref{Fig:sep_ber_coded_SE15_all_SEFDMIM}, the data BER of four SEFDM-IM systems clip in the low \(E_{b}/N_{0}\) regime, which results in the sudden drop in average BER. We find that SEFDM-IM-2 has the lowest data BER because of its inherent repetition coding in pattern-2. In addition, it is observed that the average error performance of the SEFDM-IM systems with QPSK modulation is dominated by the decision errors in index bits, where the error cliff appears. Since the transmission power of activated subcarriers is proportional to the value of \(K/K_A\) \cite{SIM_will_it_work}, the traditional SEFDM-IM-Tra system with a higher \(K/K_A\) value leads to an increased minimum Euclidean distance between different subcarrier patterns. Moreover, the single activated subcarrier in each activation pattern does not overlap for SEFDM-IM-Tra with \(K_A=1\), while in the proposed designs activation pattern-1, pattern-2 and pattern-4 activate subcarriers on repeated locations, i.e., the first and the third subcarrier locations as shown in Table \ref{tab:412_SEFDMIM}. Therefore, the bit errors resulting from erroneous detection of subcarriers' state are less likely to occur in SEFDM-IM-Tra, which leads to the best BER performance.

In Fig. \ref{Fig:sep_ber_coded_SE15}, we further compare the error performance of SEFDM-IM-2 and SEFDM-IM-Tra with classical OFDM-IM systems. Benefiting from orthogonality between subcarriers, the OFDM-IM system has superior performance in recovering QPSK data symbols, which leads to the lowest data BER. However, the OFDM-IM system suffers from higher index BER compared to that of the SEFDM-IM-Tra since OFDM-IM has a lower \(K/K_A\) according to the conclusion from \cite{SIM_will_it_work}.

The PAPR comparisons of systems at 0.75 bit/s/Hz are provided in Fig. \ref{Fig:papr_SE15}. The traditional SEFDM-IM-Tra system has the lowest PAPR, and it achieves 1.75 dB performance gain at the CCDF of \(10^{-2}\) compared to the OFDM-IM system. Meanwhile, the three proposed SEFDM-IM-1\&2\&3 systems exhibit close PAPR performance and achieve 0.6 dB gain over the OFDM-IM counterpart. Therefore, Fig. \ref{Fig:papr_SE15} reveals that PAPR becomes better when the value of \(K/K_A\) increases.

\begin{figure}[]
\begin{center}
\includegraphics[width=\columnwidth]{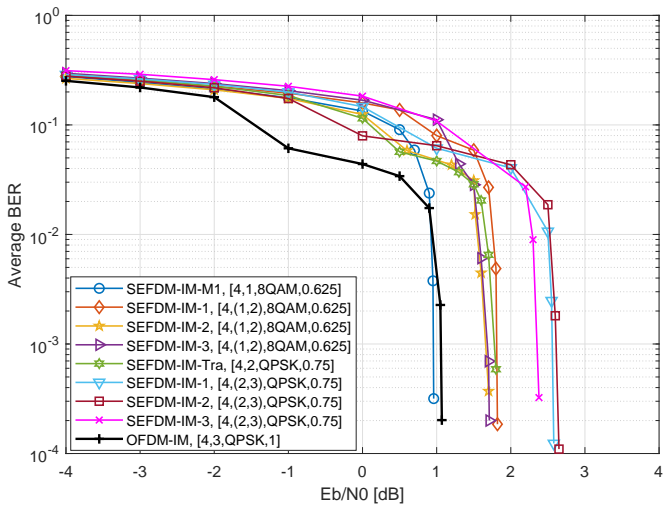}
\end{center}
\caption{Error performance of the coded SEFDM-IM and OFDM-IM systems with the spectral efficiency of 1 bit/s/Hz.}
\label{Fig:coded_SE2}
\end{figure}

\begin{figure}[]
\begin{center}
\includegraphics[width=\columnwidth]{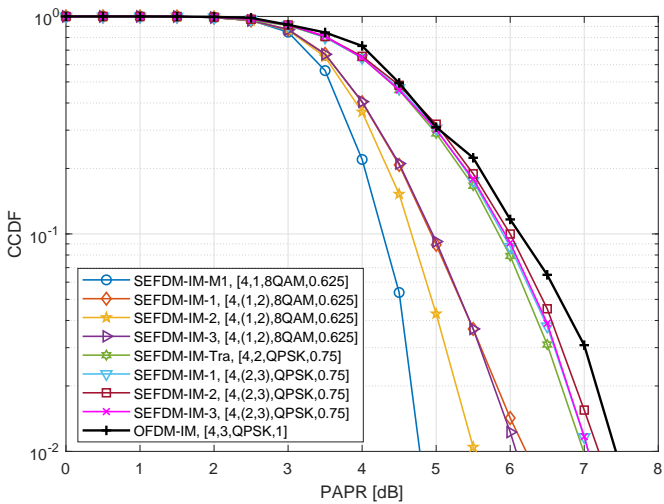}
\end{center}
\caption{CCDF of the PAPR of SEFDM-IM and OFDM-IM systems with the spectral efficiency of 1 bit/s/Hz.}
\label{Fig:papr_SE2}
\end{figure}

We then consider SEFDM-IM systems with the spectral efficiency of 2 bit/s/Hz, which becomes 1 bit/s/Hz after coding. In general, we have two ways to increase spectral efficiency of index-modulated systems: first, increase the modulation cardinality, and second, increase the number of activated subcarriers. For SEFDM-IM systems, another adjustable parameter is the level of bandwidth compression specified by \(\alpha\). Therefore, SEFDM-IM systems are more flexible in terms of achieving any desired spectral efficiency, compared to OFDM-IM systems. In our case, the increase in spectral efficiency from 0.75 to 1 bit/s/Hz is achieved by either increasing \(M_A\) to 8 or increasing \(K_A\) to 2, and \(\alpha\) is adjusted accordingly. In Fig. \ref{Fig:coded_SE2}, we observe that the SEFDM-IM systems with the increased modulation cardinality perform better than their counterparts with a higher number of activated subcarriers. This is because the dominant errors result from the decision errors in index bits, and systems with a higher value of \(K/K_A\) perform better on detecting activation patterns. A coded OFDM-IM system with \( [K,K_A]=[4,3]\) and QPSK modulation at 1 bit/s/Hz is considered. It has close BER performance to the SEFDM-IM-M1 system with \([K,K_A]=[4,1]\) and 8QAM modulation, while it suffers from 2.5 dB performance loss in terms of the PAPR at the CCDF of \(10^{-2}\), as seen in Fig. \ref{Fig:papr_SE2}.

\begin{figure}[]
\begin{center}
\includegraphics[width=\columnwidth]{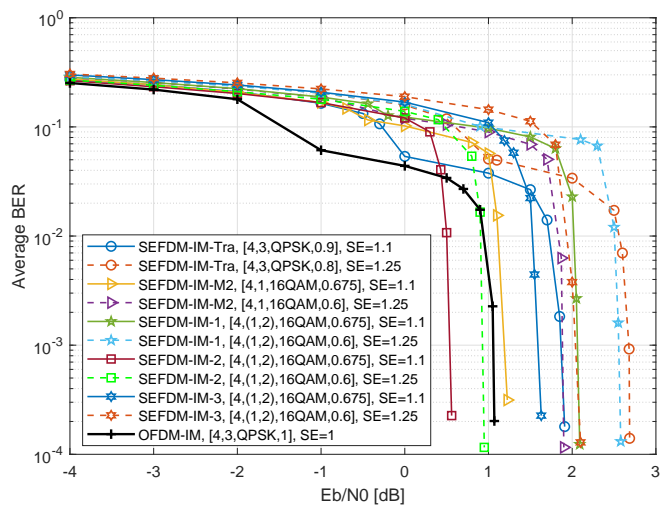}
\end{center}
\caption{Error performance of the coded SEFDM-IM systems with the spectral efficiency of 1.1 and 1.25 bit/s/Hz.}
\label{Fig:coded_SE22_25}
\end{figure}

\begin{figure}[]
\begin{center}
\includegraphics[width=\columnwidth]{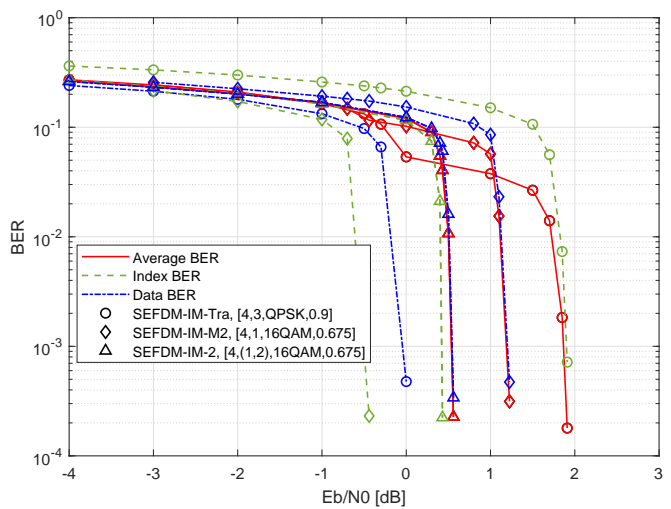}
\end{center}
\caption{Error performance of coded SEFDM-IM systems with the spectral efficiency of 1.1 bit/s/Hz in terms of index BER, data BER and average BER.}
\label{Fig:sep_ber_coded_SE22}
\end{figure}

It is observed that SEFDM-IM systems perform better with high \(K/K_A\) in terms of both BER and PAPR performance, and hence we only consider systems with \([K,K_A]=[4,1]\) and \([K,K_A]=[4,(1,2)]\) for higher spectral efficiency. 
In Fig. \ref{Fig:coded_SE22_25}, we compare BER performance of coded SEFDM-IM systems at 1.1 and 1.25 bit/s/Hz. For the spectral efficiency of 1.1 bit/s/Hz, the traditional SEFDM-IM-Tra system with \([K,K_A]=[4,3]\) and QPSK modulation operates with \(\alpha=0.9\), while the SEFDM-IM-M2 and SEFDM-IM-1\&2\&3 systems require an increased level of bandwidth compression, i.e., \(\alpha\) is reduced to 0.675. 
Compared with the traditional system, the proposed SEFDM-IM-2, SEFDM-IM-M2 and SEFDM-IM-3 systems obtain 1.3, 0.7 and 0.3 dB better BER performance, respectively. Furthermore, the best-performing SEFDM-IM-2 system achieves both 0.4 dB power gain and 10\% bandwidth saving compared to the classical OFDM-IM system at 1 bit/s/Hz. 
In Fig. \ref{Fig:sep_ber_coded_SE22}, the BER corresponding to index bits and data bits are presented to elaborate why the SEFDM-IM-2 system exhibits a BER performance advantage over the rest SEFDM-IM systems for high spectral efficiency. This is because when a high modulation cardinality is deployed, i.e., 16QAM in this paper, the average BER is dominated by the data BER, and SEFDM-IM-2 with inherent repetition coding has the best data BER performance among SEFDM-IM systems. By contrast, the traditional SEFDM-IM-Tra system with \([K,K_A]=[4,3]\) suffers from high index BER, resulting in the worst average BER.
Similar results are observed for SEFDM-IM systems at 1.25 bit/s/Hz, whose spectral efficiency is achieved by further reducing the values of \(\alpha\). In this case, BER performance of all SEFDM-IM systems are degraded due to the increased level of ICI. The performance gap between the SEFDM-IM-2 system and the traditional SEFDM-IM-Tra system is enlarged to 1.7 dB. In addition, the SEFDM-IM-2 system exhibits close BER performance to the classical OFDM-IM system while obtaining 25\% higher spectral efficiency.

\begin{figure}[]
\begin{center}
\includegraphics[width=\columnwidth]{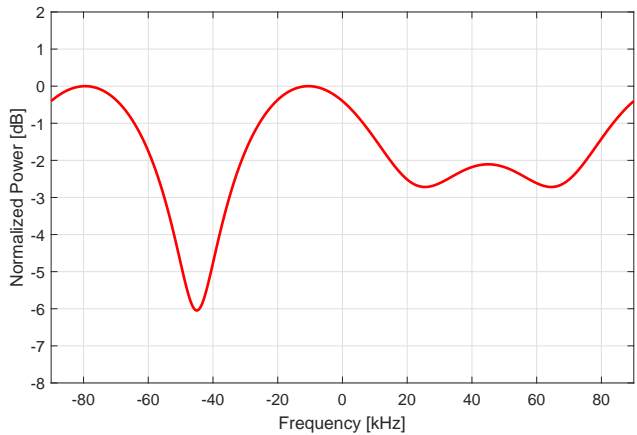}
\end{center}
\caption{ Frequency response for the static frequency selective channel model, where a deep frequency notch and two shallow frequency notches are intentionally designed.}
\label{Fig:frequency_response_channel}
\end{figure}

\begin{figure}[]
\begin{center}
\includegraphics[width=\columnwidth]{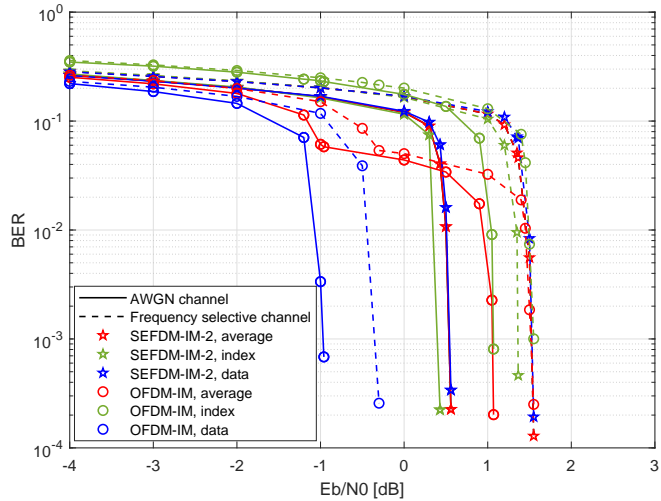}
\end{center}
\caption{Error performance of the coded SEFDM-IM-2 system at 1.1 bit/s/Hz and the OFDM-IM system at 1 bit/s/Hz with and without frequency selective channel.}
\label{Fig:fading_channel_ber}
\end{figure}

It should be noted that the systems in this paper are considered to follow NB-IoT configurations where \(N=12\) subcarriers are used. In this application area, signals normally experience flat fading given the narrow bandwidth of 180 kHz, which is the main reason why AWGN channel is normally assumed. To further evaluate the robustness of our proposals against frequency selectivity, we define a three-path static frequency selective channel given by \(h(t)=0.9137\delta(t)+0.3179\delta(t-2T_s )-0.2532e^{\frac{j\pi}{2}} \delta(t-3T_s )\), where $T_s$ indicates the time duration of one sample. Therefore, the delay spread is 16.67 \(\mu\)s considering the 180 kHz NB-IoT signal configurations. The channel frequency response is illustrated in Fig.  \ref{Fig:frequency_response_channel}, in which a deep frequency notch and two shallow frequency notches are intentionally designed to test the robustness of our proposals. For simplicity, we only show the BER performance of the best-performing SEFDM-IM-2 system discussed in Fig. \ref{Fig:coded_SE22_25} and compare it with OFDM-IM. In Fig. \ref{Fig:fading_channel_ber}, we observe that SEFDM-IM-2 and OFDM-IM experience 1 dB and 0.5 dB performance loss in terms of the average BER, respectively, when the frequency selective channel is applied. More specifically, the power gain of the index part of SEFDM-IM-2 to that of OFDM-IM decreases from 0.64 dB to 0.19 dB. Meanwhile, the power penalty of the data part of SEFDM-IM-2 to that of OFDM-IM increases by 0.3 dB. In general, both index bits and data bits in SEFDM-IM-2 are less robust to frequency selectivity when compared to those in OFDM-IM, although SEFDM-IM-2 maintains its BER advantage in index BER. This may be explained by the fact that OFDM-IM with \([K,K_A]=[4,3]\) has more subcarriers activated and therefore it is more robust to frequency selectivity when compared with SEFDM-IM-2 with \([K,K_A]=[4,(1,2)]\).

\begin{figure}[]
\begin{center}
\includegraphics[width=\columnwidth]{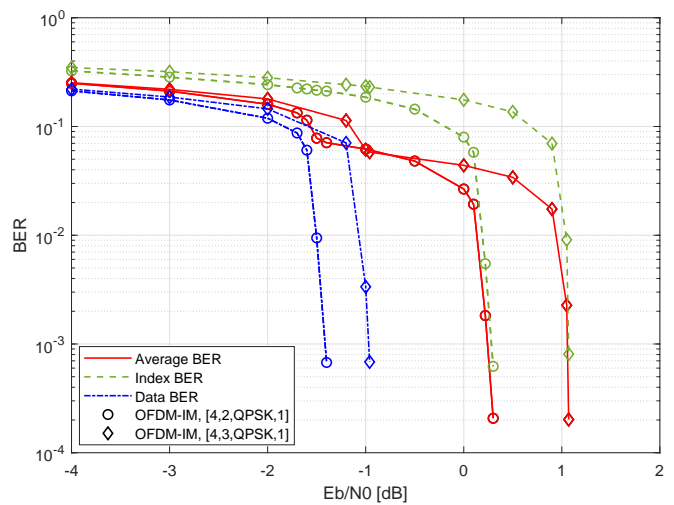}
\end{center}
\caption{Error performance of OFDM-IM systems with the different number of activated subcarriers in one subblock.}
\label{Fig:influence_of_no_of_activated_subcarriers}
\end{figure}

\begin{figure}[]
\begin{center}
\includegraphics[width=\columnwidth]{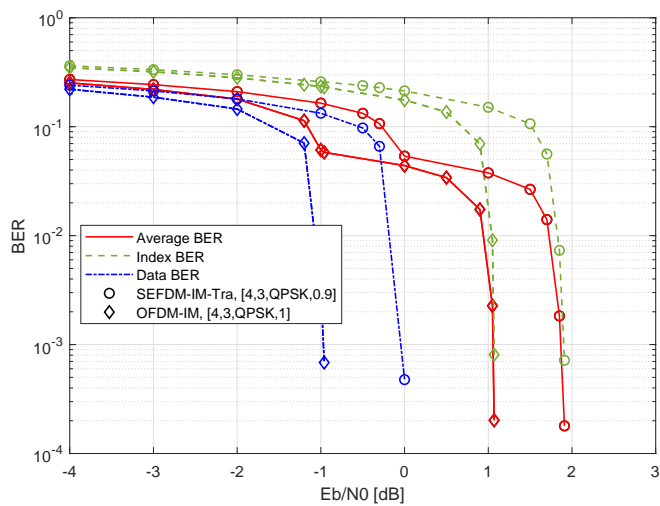}
\end{center}
\caption{Error performance of SEFDM-IM and OFDM-IM systems with different levels of bandwidth compression.}
\label{Fig:influence_of_BW_compression}
\end{figure}

\begin{figure}[]
\begin{center}
\includegraphics[width=\columnwidth]{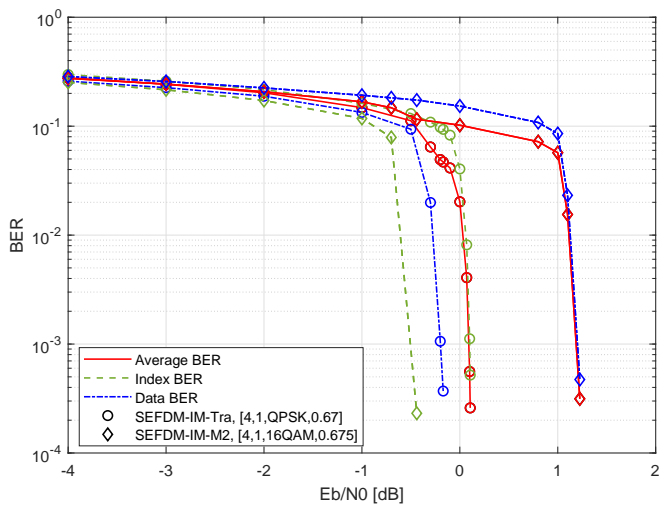}
\end{center}
\caption{Error performance of SEFDM-IM systems with different modulation schemes.}
\label{Fig:influence_of_modulation_schemes}
\end{figure}

The average BER performance is dependant on the relative performance of index BER and data BER, which is subject to the values of \(K/K_A\), \(\alpha\) and \(M_A\). In other words, the optimal SEFDM-IM design is subject to the dominant BER factor, i.e., index BER or data BER, and hence associated with the target spectral efficiency. In order to show the impact of each factor on error performance, six systems that have been discussed above are further compared. In Fig. \ref{Fig:influence_of_no_of_activated_subcarriers}, we observe that as \(K_A\) of OFDM-IM increases from 2 to 3, both index BER and data BER increase, which leads to the increase in average BER. This is because the transmission power of activated subcarriers decreases as the value \(K/K_A\) decreases. In Fig. \ref{Fig:influence_of_BW_compression}, the impact of the bandwidth compression level is presented, where 1 dB performance loss is observed on the index, data as well as average BER when a 10\% bandwidth compression is performed. In terms of the impact of \(M_A\), the data BER is degraded by 1.4 dB when QPSK modulation is replaced with 16QAM as shown in Fig. \ref{Fig:influence_of_modulation_schemes}. Consequently, the error performance of SEFDM-IM-Tra is dominated by the index part, while that of SEFDM-IM-M2 is dominated by the data part. In general, the index BER and data BER are jointly affected by the values of \(K/K_A\), \(M_A\) and \(\alpha\). It is inferred that both index BER and data BER increase with the decrease in \(K/K_A\) or \(\alpha\) values, and data BER increases with the increase in \(M_A\).

\begin{figure}[]
\begin{center}
\includegraphics[width=\columnwidth]{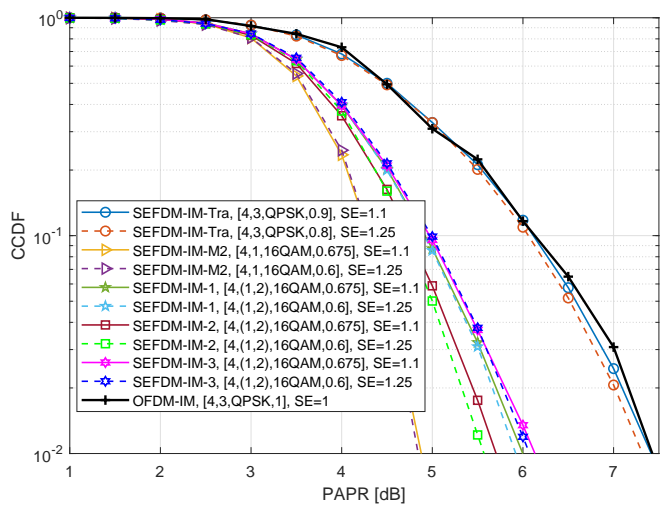}
\end{center}
\caption{CCDF of the PAPR of SEFDM-IM systems with the spectral efficiency of 1.1 and 1.25 bit/s/Hz.}
\label{Fig:papr_SE22_25}
\end{figure}

The PAPR performance of the systems with 1.1 and 1.25 bit/s/Hz is shown in Fig. \ref{Fig:papr_SE22_25}. As expected, systems with \([K,K_A]=[4,3]\) exhibit higher PAPR than those with \([K,K_A]=[4,(1,2)]\), and the SEFDM-IM-M2 systems with \([K,K_A]=[4,1]\) achieve the lowest PAPR. For systems with the same configurations except for the value of \(\alpha\), PAPR slightly decreases when \(\alpha\) decreases from 0.9 to 0.8 and from 0.675 to 0.6. This demonstrates the advantage of SEFDM-IM systems in terms of PAPR reduction.

\section{Computational Complexity Analysis}\label{section5}

In this section, we compare the computational complexity of the investigated systems. Since all of them need to perform standard LDPC decoding, we can only consider the computational complexity of LLR calculations. According to \eqref{eq:LLR_index_3} for index bits and \eqref{eq:LLR_data_2} for data bits, the computational complexity of LLR calculations is dominated by \(\Psi\left ( {I}^{g},{P}^{g} \right )\) calculations and therefore we use the number of \(\Psi\left ( {I}^{g},{P}^{g} \right )\) calculations required per coded bit as the metric for complexity comparisons.
The calculations of \(\Psi\left ( {I}^{g},{P}^{g} \right )\) need to be performed for all possible combinations of valid activation patterns and data symbols, given by
\begin{eqnarray}\label{eq:computational_complexity_calculations}
{\forall{{I}}^{g}\in \mathcal{I},\forall{{{P}}^{g}\in \Upsilon^{K_A}}:{\Psi}\left ( {I}^{g},{P}^{g} \right ) ,}
\end{eqnarray} 
where \(g=1,2,...,G\), and \(\forall\) denotes the operation that loops through all elements in a given set. Since the calculation results remain the same for all coded bits in one subblock, \eqref{eq:computational_complexity_calculations} only needs to be performed once per subblock. In any index-modulated systems with the traditional subcarrier pattern design, both \(M\) and \(K_A\) have a single value. As a result, the computational complexity per coded bit in terms of the number of \(\Psi\left ( {I}^{g},{P}^{g} \right )\) calculations is given by  
\begin{eqnarray}\label{eq:computational_complexity_traditional}
{{\Theta}_{tra}=\frac{1}{L}U\left (M\right )^{{K_A}},}
\end{eqnarray}  
where the subscript \(tra\) stands for the traditional subcarrier pattern design.
By contrast, SEFDM-IM-1\&2\&3 with the proposed subcarrier pattern design has two values of \(K_A\) and multiple modulation cardinalities because of pattern-2. Since we deliberately set a constant number of data bits transmitted per subblock regardless of activation patterns, the computational complexity of pattern-2 is the same as that of other patterns. Therefore, the total number of \(\Psi\left ( {I}^{g},{P}^{g} \right )\) calculations required by SEFDM-IM-1\&2\&3 is calculated same as the traditional counterpart. Hence, the computational complexity of obtaining the LLR value for a coded bit transmitted in SEFDM-IM-1\&2\&3 is given by
\begin{eqnarray}\label{eq:computational_complexity}
{{\Theta}_{pro}=\frac{1}{L}U\left (M'  \right )^{{K_A}'},}
\end{eqnarray}  
where the subscript \(pro\) represents the proposed subcarrier pattern design, and \({M}'\) and \({K_A}'\) are the number of activated subcarriers per subblock and the modulation cardinality used in pattern-1, respectively. Taking the configuration of \([K,K_A]=[4,(1,2)]\) in Table \ref{tab:412_SEFDMIM} as an example, ${M}'$ is replaced by $M_A$, and ${K_A}'$ is assigned with a value of 1 regardless of ${K_A}=(1,2)$. The computational complexity of OFDM is not considered here since OFDM detectors are based on simple subcarrier-based detection as opposed to the subblock-based detection required for the IM-based cases.

\begin{table}[t]
\caption{\\Computational complexity of index-modulated systems.}
\centering
\begin{tabular}{|c|c|c|}
\hline&&\\[-0.65em]
\textit{\textbf{Scheme}}                                  &\textit{\textbf{\begin{tabular}[c]{@{}c@{}}Modulation scheme\\in pattern-1\&3\&4 \end{tabular}}} & \textit{\textbf{\begin{tabular}[c]{@{}c@{}}Computational\\ complexity\end{tabular}}}\\ 
\hline\hline &&\\[-0.65em]
\multirow{3}{*}{\begin{tabular}[c]{@{}c@{}}SEFDM-IM with \\ \([K,K_A]=[4,1]\),\\[0.5ex] \([K,K_A]=[4,(1,2)]\)\end{tabular}} & QPSK     & 4 \\[0.5ex]   \cline{2-3}\cline{2-3} &&\\[-0.65em]  & 8QAM & 7 \\ [0.5ex]  \cline{2-3}\cline{2-3} &&\\[-0.65em]
 & 16QAM    & 11   \\[0.5ex]  \hline &&\\[-0.65em]
\begin{tabular}[c]{@{}c@{}}SEFDM-IM with\\ \([K,K_A]=[4,2]\),\\[0.5ex] \([K,K_A]=[4,(2,3)]\);\\[0.8ex] OFDM-IM with\\ \([K,K_A]=[4,2]\)\end{tabular} 
& QPSK  & 11     \\ [1.1ex] \hline &&\\[-0.65em]
\begin{tabular}[c]{@{}c@{}}SEFDM-IM with\\ \([K,K_A]=[4,3]\);\\ [0.8ex] OFDM-IM with\\ \([K,K_A]=[4,3]\)\end{tabular}  & QPSK    & 32  \\ 
\hline
\end{tabular}
\label{tab:computational_complexity}
\end{table}

The computational complexity of all investigated systems with the LLR calculator is provided in Table \ref{tab:computational_complexity}\footnote{Modulation schemes for pattern-1\&3\&4 are fixed. Pattern-2 could employ mixed modulation schemes but has the same computational complexity with pattern-1\&3\&4.}. It is observed that the computational complexity increases exponentially upon the linearly increased \(K_A\) value, such that a system with a low \(K_A\) value achieves reasonable computational complexity despite a high modulation cardinality. Therefore, results in Table \ref{tab:computational_complexity} reveal that the proposed SEFDM-IM systems with \([K,K_A]=[4,(1,2)]\) achieve reduced computational complexity, compared to traditional SEFDM-IM-Tra with \([K,K_A]=[4,2]\) and \([K,K_A]=[4,3]\) under equivalent spectral efficiency. 
Furthermore, we find that classical OFDM-IM systems typically suffer from high computational complexity. This is because they are required to transmit more bits to achieve the target spectral efficiency compared to their SEFDM-IM counterparts, due to the constraint on subcarrier spacing.

\section{Conclusion}\label{section6}

In this paper, we have proposed a novel index modulation pattern design principle for SEFDM-IM systems based on keeping the last subcarrier in each subblock unused, thereby improving signal quality, since ICI levels are dependant on  locations of activated subcarriers. The deletion of the last subcarrier is compensated by varying the number of activated subcarriers per subblock. Following this design principle, we have developed three SEFDM-IM activation schemes termed SEFDM-IM-1\&2\&3 and compared their system performance with that of classical OFDM-IM and traditional SEFDM-IM. In addition to BPSK and QPSK that were investigated in previous work, we have explored higher-order modulation schemes 8QAM and 16QAM with both traditional and proposed activation patterns. Results have shown that the newly proposed schemes outperform OFDM-IM and other SEFDM-IM in terms of BER, PAPR and computational complexity, when spectral efficiency is 1 bit/s/Hz and higher. Furthermore, we have found that SEFDM-IM systems with a higher modulation cardinality outperform those with a higher number of activated subcarriers at equivalent spectral efficiency. Therefore, this work provides a useful design principle for coded SEFDM-IM by tuning subcarrier activation patterns, number of activated subcarriers, bandwidth compression factors and modulation schemes.

\bibliographystyle{IEEEtran}
\bibliography{Paper-waveform-IM-2022}

\end{document}